\documentclass[9pt,a4paper,twoside]{rho-class/rho}
\usepackage[english]{babel}

\setbool{rho-abstract}{true} 
\setbool{corres-info}{true} 

\title{Sparsity, Regularization and Causality in Agricultural Yield: The Case of Paddy Rice in Peru}

\author[1,a]{Rita Rocio Guzman-Lopez}
\author[1,b]{Luis Huamanchumo}
\author[2]{Kevin Fernandez}
\author[1]{Oscar Cutipa-Luque}
\author[1]{Yhon Tiahuallpa}
\author[3,c]{Helder Rojas}

\affil[1]{Professional School of Statistical Engineering, National University of Engineering, Peru}
\affil[2]{Institute of Geosciences, University of São Paulo, Brazil}
\affil[3]{Department of Mathematics, Imperial College London, UK}
\affil[a]{Email: rrguzman@uni.edu.pe}
\affil[b]{Email: lhuamanchumo@uni.edu.pe}
\affil[c]{Email: h.rojas-molina23@imperial.ac.uk}

\dates{September 20, 2024}

\leadauthor{Guzman, R. et al.}
\footinfo{}
\smalltitle{FIEECS, UNI}
\institution{National University of Engineering}
\theday{Sept, 2024} 

\begin{abstract}

This study introduces a novel approach that integrates agricultural census data with remotely sensed time series to develop precise predictive models for paddy rice yield across various regions of Peru. By utilizing sparse regression and Elastic-Net regularization techniques, the study identifies causal relationships between key remotely sensed variables—such as NDVI, precipitation, and temperature—and agricultural yield. To further enhance prediction accuracy, the first- and second-order dynamic transformations (velocity and acceleration) of these variables are applied, capturing non-linear patterns and delayed effects on yield. The findings highlight the improved predictive performance when combining regularization techniques with climatic and geospatial variables, enabling more precise forecasts of yield variability. The results confirm the existence of causal relationships in the Granger sense, emphasizing the value of this methodology for strategic agricultural management. This contributes to more efficient and sustainable production in paddy rice cultivation.

\end{abstract}


\begin{document}
	
    \maketitle


\section{Introduction}\label{sec:1}

\rhostart{T}oday, precision agriculture is undergoing rapid transformation due to the integration of advanced technologies such as pattern recognition, machine learning, and the use of remotely sensed data and imagery \cite{liakos2018machine,kussul2017deep}. These innovations have drastically improved farmers' ability to forecast agricultural yields with unprecedented accuracy. By analyzing large volumes of data and detecting hidden patterns, these technologies enable the prediction of crop yields, identification of potential problems, and optimization of resource use, including water, fertilizers, and pesticides. However, in the regional context—particularly in Peru—the application of these technologies for agricultural production forecasting remains underexplored, leaving substantial potential yet to be tapped \cite{briceno2019deforestacion, fernandezanalisis}.

Therefore, in this work, we are interested in studying how machine learning techniques, combined with climatological and geospatial data, remote sensing data, and imagery, can be used to improve the predictive capacity of agricultural yields. In particular, we aim to investigate the identification of causal relationships between remote sensing variables, such as NDVI, precipitation, and temperature, and agricultural yields of certain crops. Additionally, we are interested in using these causal relationships to build simple and parsimonious machine learning models that accurately forecast agricultural yields. To this end, we focus on rice crop yield data in Peru as a case study for our proposed techniques and methodologies.

It is important to mention that neither the choice of the agricultural product nor the specific geographical area limits the scope of our conclusions and results. However, it is also important to highlight that, despite the relevance of rice in Peruvian agriculture, there is a scarcity of local research specifically addressing rice yield forecasting using advanced techniques. Therefore, this study specifically aims to develop and investigate how the use of sparsity, regularization, and machine learning techniques, combined with remote sensing variables, can positively influence the accuracy of agricultural yield forecasts, contributing to this sector and to the development of innovative methodologies that allow for yield prediction, production optimization, and the sustainability of this crop.

In this context, several advanced methodologies employed in this work are highlighted. Among these are techniques for extracting remote sensing data and integrating them with the National Agricultural Survey (ENA), both of which significantly influence crop yield. Notably, the regression model with Elastic-Net regularization stands out, offering enhanced flexibility by combining the penalties associated with two standard rules with desirable properties. This approach achieves a balance between variable selection and parameter regulation, which was used to identify causal relationships between remote sensing variables and agricultural yield \cite{zou2005regularization}. Since our goal is to identify parsimonious causal relationships between variables, we will employ sparsity-inducing techniques that align with qualitative field criteria. Additionally, Generalized Additive Models (GAM) will be applied to capture nonlinear relationships between predictor variables and agricultural yield. Finally, the XGBoost model will be implemented, a highly flexible machine learning technique for agricultural yield prediction. XGBoost is capable of handling large datasets and capturing complex interactions between variables, significantly improving prediction accuracy. Therefore, these last two models will be employed to obtain agricultural yield forecasts based on the previously identified causal relationships, and their results will be compared with those obtained from the Elastic-Net regularization regression model.

\subsection{Outline}
The article is organized as follows. In Section \boldref{sec:2}, we describe the study area and explain how the dataset was structured and obtained, including the extraction and processing of remote sensing data and its integration with the Peruvian National Agrarian Survey (ENA). We also provide a detailed explanation of the data preprocessing and modeling phase, where we applied techniques such as regression with Elastic-Net regularization, Generalized Additive Models (GAM), and the XGBoost model. In Section \boldref{sec:3}, we present and discuss the results obtained from these models, analyzing their performance and how causal relationships between remote sensing variables and agricultural yield were identified. In Section \boldref{sec:4}, we present the conclusions of the study, emphasizing the main findings and demonstrating how the use of sparsity, regularization, and machine learning techniques, combined with remote sensing variables, can enhance the prediction of agricultural yield. Finally, in Section \boldref{sec:5}, we outline the specific contributions made by the authors to this work.
\section{Materials and methods}\label{sec:2}
    
\subsection{Area of study}

As mentioned earlier, we use paddy rice production in Peru as a case study for our methodologies. Several regions of Peru were selected as study areas, focusing on those with the highest concentration of paddy rice production. The primary source of information for this analysis is the agricultural censuses conducted in Peru between 2015 and 2018, which provide detailed and up-to-date data on farming practices and characteristics\footnote{\url{https://www.datosabiertos.gob.pe/search/type/dataset?query=encuesta+nacional+agropecuaria&sort_by=changed&sort_order=DESC}}.

To ensure proper organization and georeferencing of the collected data, a specific coding system was developed. This system allows for the identification of the location of each study area, facilitating both spatial and temporal analysis. The coding system includes the department code (CCDD), which identifies the administrative region; the province code (CCPP), which specifies the subdivision within the department; the district code (CCDI), which details the local subdivision; and the cluster code, which groups smaller areas or sampling units within a district. Additionally, latitude and longitude coordinates were incorporated, as demonstrated in Table \boldref{tab1}, which illustrates the implementation of this coding system in the project. This highlights its applicability in identifying and monitoring paddy rice production areas over time.

\begin{table}[ht]
    \centering
    \scalebox{0.85}{%
    \begin{tabular}{cccccccc}
        \hline
        \textbf{ID} & \textbf{YEAR} & \textbf{CCDD} & \textbf{CCPP} & \textbf{CCDI} & \textbf{CONGL} & \textbf{LAT} & \textbf{LONG} \\
        \hline
        1 & 2018 & 1 & 2 & 3 & 5198 & -5.676 & -78.438 \\
        2 & 2018 & 1 & 7 & 2 & 5318 & -5.806 & -78.219 \\
        3 & 2018 & 1 & 7 & 2 & 5318 & -5.808 & -78.219 \\
        4 & 2017 & 2 & 18 & 1 & 8260 & -9.006 & -78.540 \\
        5 & 2016 & 2 & 18 & 1 & 8250 & -8.955 & -78.580 \\
        6 & 2017 & 20 & 1 & 9 & 1132 & -5.372 & -80.707 \\
        \hline
    \end{tabular}%
    }
    \caption{The table displays the location variables as follows: CCDD indicates the administrative region; CCPP specifies the subdivision within the department; CCDI details the local subdivision within a province; CONGL groups sampling areas within a district. LAT and LONG represent the precise geographic coordinates of the paddy rice crops.}
    \label{tab1}
\end{table}
\subsection{Remote sensing data}

\begin{figure}[h!]
        \centering
        \includegraphics[width=0.9\linewidth]{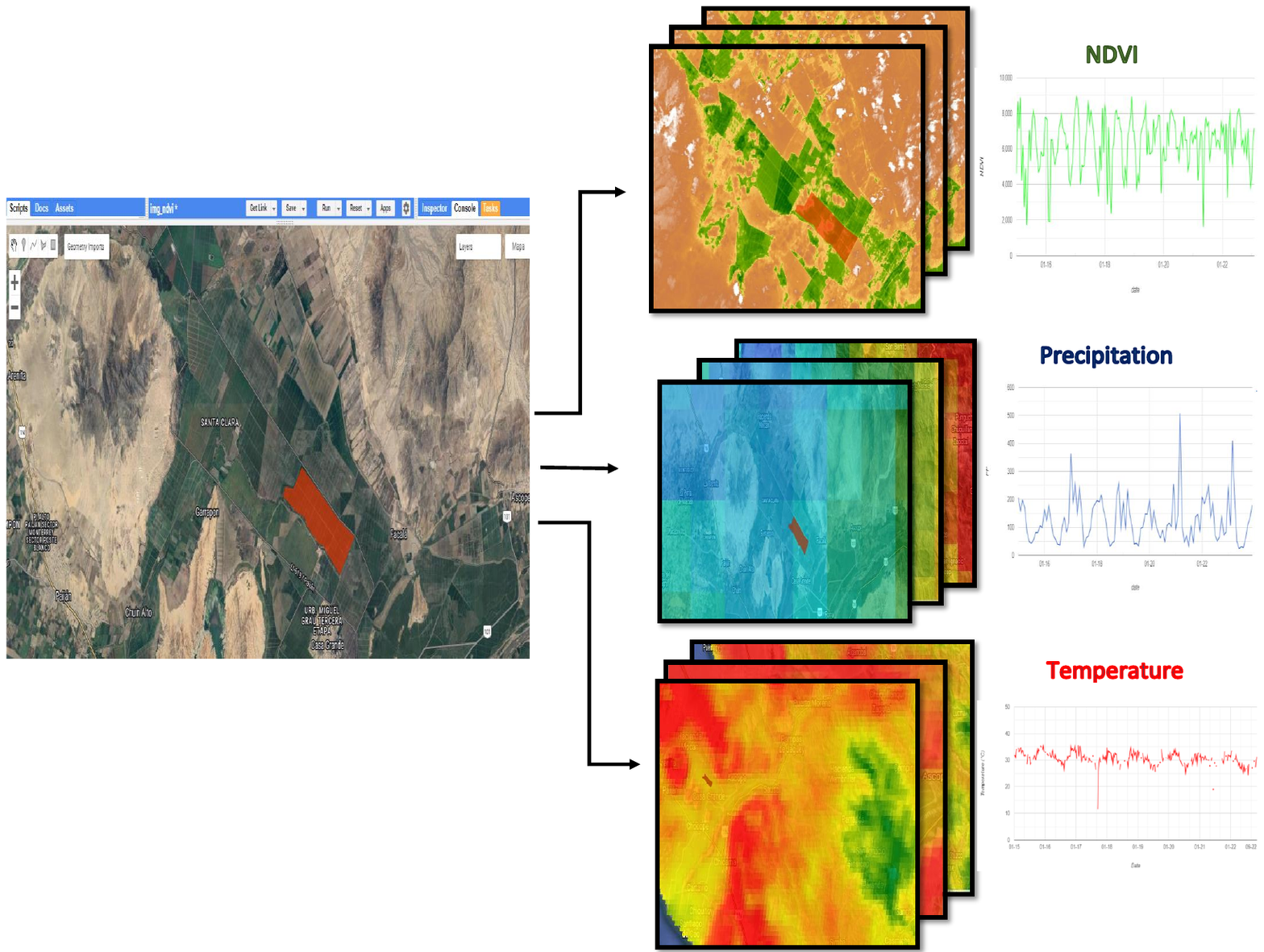}
        \caption{The graphic illustrates the process of extracting remotely sensed time series using Google Earth Engine. On the right, three images are shown, each color-coded to represent a different type of series: NDVI, precipitation, and temperature.}
        \label{sens}
    \end{figure}

The satellite images used in this study were obtained through remote sensing and processed using open-access tools like Google Earth Engine (GEE)\footnote{\url{https://code.earthengine.google.com/}}. On this platform, radiometric and atmospheric corrections were applied, and cloud-induced variability was addressed to ensure high-resolution images\cite{ambrosio2002correccion, saunders1988improved}. These images are crucial for generating accurate information from relevant indices, as shown in Figure \boldref{sens}.

\begin{figure}[h!]
    \centering
    \includegraphics[width=0.95\linewidth]{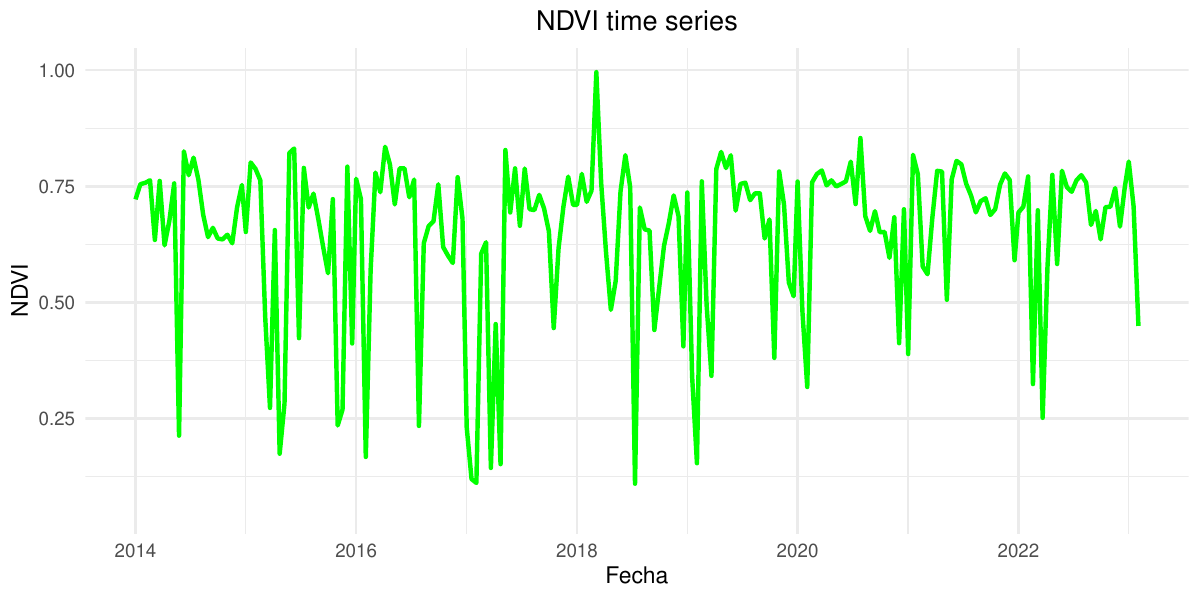}
    \caption{The graph presents a time series of the Normalized Difference Vegetation Index (NDVI) with a 16-day frequency, obtained using the MOD13Q1 product from the MODIS satellite, with a spatial resolution of 250 meters.}
    \label{ser-ndvi}
\end{figure}

Key remote sensing variables that significantly influence crop development and yield were identified and selected. The first of these is the Normalized Difference Vegetation Index (NDVI), a crucial metric for assessing vegetation health \cite{yengoh2015use}, which is calculated as follows:

\begin{equation}\label{NDVI}
  NDVI = \frac{\rho NIR - \rho RED}{\rho NIR + \rho RED}    \\ , 
\end{equation}

where NIR is the near infrared reflectance band and RED is the red band. The data for this variable was obtained through the MOD13Q1 sensor, which generates NDVI time series with a frequency of 16 days \cite{volante2015expansion}. This sensor allows categorizing land surface properties and biological processes, as well as primary production and land cover changes. The sampled NDVI series is shown in Figure \boldref{ser-ndvi}.

\begin{figure}[h!]
    \centering
    \includegraphics[width=0.95\linewidth]{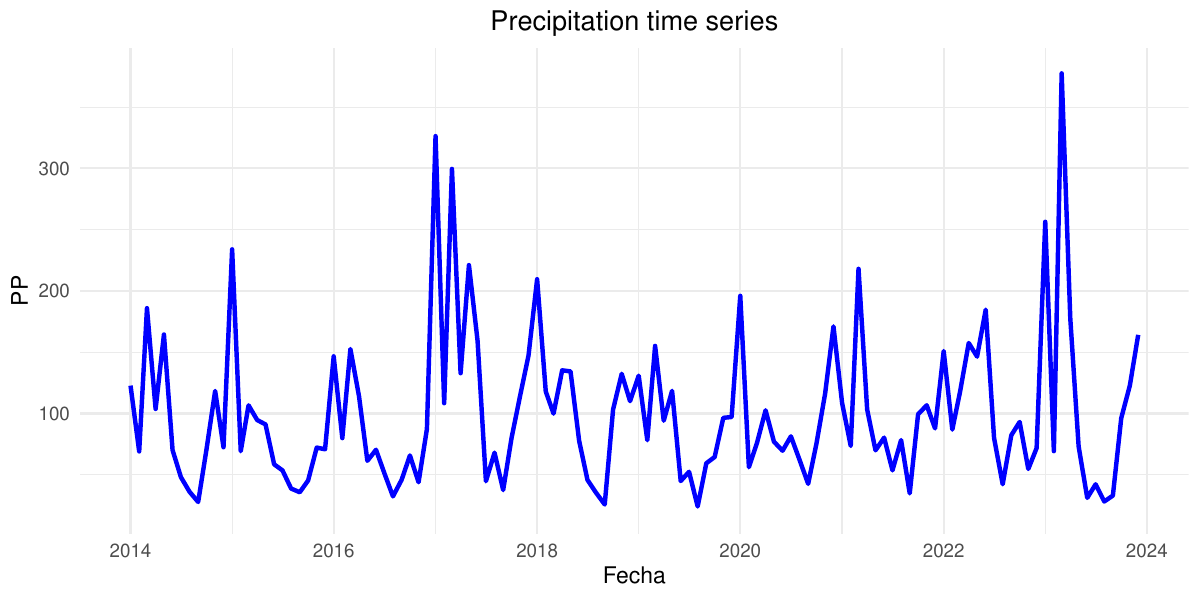}
    \caption{The graph presents a time series of monthly precipitation, measured in millimeters, using the CHIRPS product from 2014 to 2023.}
    \label{ser-prec}
\end{figure}

The second variable is Precipitation (PREC), which directly affects soil moisture and is therefore a critical factor for crop growth \cite{soto2002instructivo}. Although the specific formula for determining precipitation using CHIRPS Pentad is not explicitly provided, the process involves combining satellite data with information from weather stations \cite{funk2014quasi}. The general procedure can be described as follows:

\begin{equation}\label{chirps}
  PREC =  f(S,T,C)  \\  ,
\end{equation}

where $PREC$ represents the estimated precipitation, $S$ corresponds to satellite data, $T$ refers to ground station data, and $C$ includes applied corrections and adjustments. This process generates precipitation time series in a gridded format, which is useful for trend analysis and seasonal drought monitoring. The sampled PREC time series is shown in Figure \boldref{ser-prec}.

\begin{figure}[h!]
    \centering
    \includegraphics[width=0.95\linewidth]{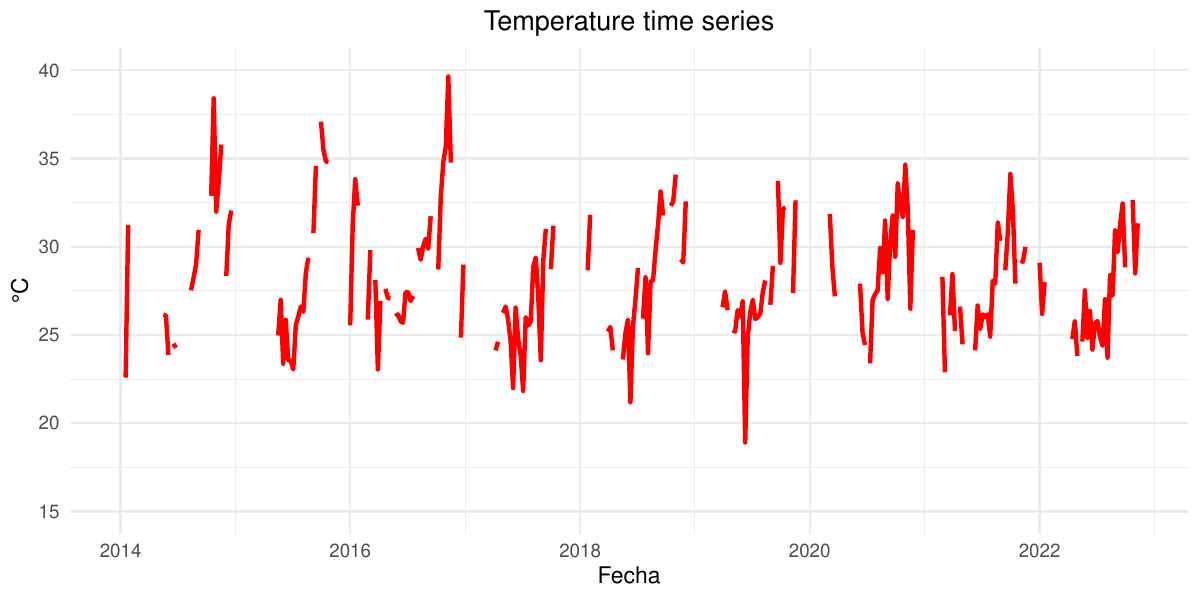}
    \caption{The graph presents a time series of land surface temperature (LST), measured in degrees Celsius on a daily basis, provided by the MOD11A1 sensor, covering the period from June 2014 to September 2023.}
    \label{ser-temp}
\end{figure}

Finally, the third remote sensing variable is Temperature (TEMP), which plays a crucial role in crop germination and development \cite{sanchez2012efecto}. This variable is obtained using the MOD11A1 sensor, and although no specific formula is provided, the general formula is as follows:
\begin{equation}\label{temp}
  TEMP = a + b(T_{31} + T_{32}) + c(T_{31} - T_{32}) + d(T_{31} - T_{32})^2 \\ ,
\end{equation}
where $LST$ represents the land surface temperature, and $T_{31}$ and $T_{32}$ are the brightness temperatures in bands 31 and 32, respectively \cite{guha2019analytical}. The coefficients $a, b, c, d$ are empirically determined and vary with atmospheric and surface conditions. To estimate land surface temperature, algorithms that combine satellite data with weather station observations are employed. The sampled TEMP time series is shown in Figure \boldref{ser-temp}.

\subsection{Data preprocessing}   
\label{Sub: Vel_Acel}

Once the remotely sensed data and relevant agricultural census data were extracted, it was crucial to ensure robust temporal consistency and uniform quality and frequency before proceeding with the analysis. To achieve this, a Spline interpolation process was implemented, allowing the establishment of a weekly frequency in the NDVI, PREC, and TEMP series \cite{martinez2009vegetacion, wongsai2017annual}. This process not only enhanced the accuracy in capturing the temporal variability of the data, but also enabled the division of the data into twelve lags, referred to as \textit{lags}, for the NDVI, Precipitation, and Temperature variables. As a result, after interpolation, the three remote sensing variables are provided as time series with a weekly frequency.

Additionally, considering the complex and highly nonlinear nature of the relationships between the remote sensing variables and agricultural yield, we chose to include both first- and second-order variations of the NDVI, Precipitation, and Temperature variables, along with their respective time lags. Due to the physical interpretation of these first- and second-order differences, we refer to these new variables as the velocities and accelerations of the remotely sensed variables. For instance, NDVI velocity—analogously defined for the Precipitation and Temperature variables—is described as the rate of change in NDVI values between consecutive periods:

\begin{equation}\label{vel_NDVI}
  VEL\_NDVI_{t} := \Delta\, NDVI_{t} =NDVI_{t} - NDVI_{t-1},  
\end{equation}

where $NDVI_{t}$ is the NDVI corresponding to week $t$. Additionally, NDVI acceleration—analogously defined for the Precipitation and Temperature variables—represents the change in NDVI velocity between consecutive periods:
\begin{equation}\label{acel_NDVI}
  ACCEL\_NDVI_{t} := \Delta^2\, NDVI_{t}= VEL\_NDVI_{t}-VEL\_NDVI_{t-1}.  
\end{equation}

Once the velocity and acceleration variables were defined for the three series, we also incorporated the lags of up to 12 weeks for these new variables into the analysis. Specifically, the following sequences of variables were considered
    \begin{align}
    &\Big\{VEL\_NDVI_{t-d}\Big\}_{d=1}^{12},\quad \Big\{ACCEL\_NDVI_{t-d}\Big\}_{d=1}^{12},\, \qquad  \\
    &\Big\{VEL\_PREC_{t-d}\Big\}_{d=1}^{12},\quad \Big\{ACCEL\_PREC_{t-d}\Big\}_{d=1}^{12},\, \qquad \\
    &\Big\{VEL\_TEMP_{t-d}\Big\}_{d=1}^{12},\quad \Big\{ACCEL\_TEMP_{t-d}\Big\}_{d=1}^{12}.\, \qquad
    \end{align}     

These newly derived variables allow us to capture higher-order dynamic relationships that would be difficult to detect using the original variables alone. An important finding in this study, as we will demonstrate, is that these new variables significantly improve the predictive capacity of the models. They enable us to incorporate dynamic effects into the relationships in a straightforward manner.

\subsection{Data Set}

The development of the dataset for this study involved several key considerations. First, we ensured that the crop under study was homogeneous, focusing on agricultural areas where only one type of crop was grown, which allowed for more accurate data collection. Additionally, we selected a transitory crop, which undergoes distinct phenological stages such as sowing, growth, and harvest. Another crucial criterion was the ability to clearly observe and distinguish the crop using satellite imagery. For these reasons, along with its social importance, we chose to study paddy rice, a crop of significant relevance in Peru that met all the above criteria.

Once the crop areas were identified, we proceeded with the extraction of their variables and characteristics, drawing from two main data sources. The first source was the National Agrarian Survey (ENA), which includes variables characterizing the use of good agricultural practices on the farms associated with the sampled crop areas. The second data source consisted of remote sensing data, extracted using open-access tools such as Google Earth Engine (GEE). For this source, algorithms were developed to obtain spatial information on NDVI, PREC, and TEMP, based on latitude, longitude, and multispectral images of the sampled plots. These series were then interpolated to obtain weekly observations, with lags of up to twelve weeks before the harvest date.

It is important to note that limitations related to inaccuracies in the geographic information system for rice cultivation provided by the ENA required significant resources for data cleaning and correction. Combined with the limited resources available for the study, this resulted in the identification of only 348 paddy rice plots across different regions of Peru. However, it is worth emphasizing that with additional resources, we could significantly expand the sample size, thereby strengthening our results and conclusions.

Finally, after completing the variable engineering process—which involved creating velocity and acceleration variables from the remote sensing data—we integrated this information with control variables extracted from the National Agrarian Survey, as shown in Table \boldref{Tabla:Variables}. This integration resulted in a comprehensive dataset of 348 records, which served as the basis for the analysis conducted in the modeling phase described in detail below.

\begin{table*}[!h]
\centering
\begin{tabular}{cp{10cm}c}
\textbf{Label} & \textbf{Description} & \textbf{Source} \\
\hline \hline
Prod\_Hect & Agricultural production of paddy rice measured in tons per hectare & ENA \\
\hline
P204\_TIPO & Type of crop. \newline 1: Homogeneous \quad 2: Heterogeneous & ENA \\
\hline
P206\_INI & Harvest start date & ENA \\
\hline
P208 & Crop management \newline 1: Homogeneous \quad 2: Associated \quad 3: Dispersed & ENA \\
\hline
P211\_1 & When planting paddy rice, did you consider the climate of the area? \newline 0: No \quad 1: Yes & ENA \\
\hline
P211\_2 & When planting paddy rice, did you consider the availability of water? \newline 0: No \quad 1: Yes & ENA \\
\hline
P211\_4 & When planting paddy rice, did you consider the type of soil? \newline 0: No \quad 1: Yes & ENA \\
\hline
P212 & The water for irrigating the crop comes from \newline 1: Rain \quad 2: River \quad 3: Spring \quad 4: Groundwater \newline 5: Reservoir \quad 6: Dam \quad 7: Other & ENA \\
\hline
P213 & The irrigation system used was \newline 1: Exudation \quad 2: Drip \quad 3: Microsprinkler \quad 4: Sprinkler \newline 5: Multi-gates \quad 6: Hoses \quad 7: Gravity \quad 8: Others & ENA \\
\hline
P214 & The seed used was \newline 1: Certified \quad 2: Non-certified & ENA \\
\hline
$NDVI_{t}$ & Normalized Difference Vegetation Index (NDVI) corresponding to the week of paddy rice harvest \( t \) & Remote Sensing \\
\hline
$PREC_{t}$ & Precipitation corresponding to the week of paddy rice harvest \( t \) & Remote Sensing \\
\hline
$TEMP_{t}$ & Temperature corresponding to the week of paddy rice harvest \( t \) & Remote Sensing \\
\hline
$\Delta\, NDVI_{t}$ & Velocity of the Normalized Difference Vegetation Index corresponding to the week of harvest \( t \) & Remote Sensing \\
\hline
$\Delta\, PREC_{t}$ & Velocity of Precipitation corresponding to the week of harvest \( t \) & Remote Sensing \\
\hline
$\Delta\, TEMP_{t}$ & Velocity of Temperature corresponding to the week of harvest \( t \) & Remote Sensing \\
\hline
$\Delta^2\, NDVI_{t}$ & Acceleration of the Normalized Difference Vegetation Index corresponding to the week of harvest \( t \) & Remote Sensing \\
\hline
$\Delta^2\, PREC_{t}$ & Acceleration of Precipitation corresponding to the week of harvest \( t \) & Remote Sensing \\
\hline
$\Delta^2\, TEMP_{t}$ & Acceleration of Temperature corresponding to the week of harvest \( t \) & Remote Sensing \\
\hline \hline
\end{tabular}
\caption{Labels, descriptions, and sources of the variables used.}
\label{Tabla:Variables}
\end{table*}

\subsection{Modeling Phase}

The final dataset considered for this study is composed of $\mathcal{D}=\{{(\boldsymbol{x}_i, y_i)}\}_{i=1}^{N}$, where $N=348$ represents the number of sampled plots. Here, the response variable $y_i \in \mathbb{R}$, labeled as Prod-Hect, denotes the agricultural yield of the crop, measured in tons per hectare, where the harvest occurred in week $T_i$. Additionally, the covariate vector $\boldsymbol{x}_i\in \mathbb{R}^{81}$ consists of two groups of variables.

The first group, denoted by $\boldsymbol{z}_i\in \mathbb{R}^{9}$, includes variables that characterize the application of good agricultural practices for crop $i$. This set of variables is defined as follows:
    \begin{equation}
      \begin{split}
        \boldsymbol{z}_i = (P204\_TIPO_i, P206\_INI_i, P208_i, P211\_1_i, \qquad\qquad \\ 
       \quad P211\_2_i, P211\_4_i, P212_i, P213_i, P213_i),
      \end{split}
    \end{equation}

where the labels are described in Table \boldref{Tabla:Variables}. It is important to note that the variables comprising $\boldsymbol{z}_i$ were extracted from the ENA, corresponding to the year immediately following the harvest week $T_i$. The second group consists of the series $VEL\_NDVI_{t}^i$, $ACCEL\_NDVI_{t}^i$, $VEL\_PREC_{t}^i$, $ACCEL\_PREC_{t}^i$, $VEL\_TEMP_{t}^i$, and $ACCEL\_TEMP_{t}^i$, where the superscript indicates that these series pertain to crop $i$. This is expressed as follows:
\begin{equation}
    \begin{split}\label{w}
    \boldsymbol{w}_i=\Bigg( \Big\{VEL\_NDVI_{T_i-d}^i\Big\}_{d=1}^{12},\, \Big\{ACCEL\_NDVI_{T_i-d}^i\Big\}_{d=1}^{12},\, \qquad  \\
    \Big\{VEL\_PREC_{T_i-d}^i\Big\}_{d=1}^{12},\, \Big\{ACCEL\_PREC_{T_i-d}^i\Big\}_{d=1}^{12},\, \qquad \\
    \Big\{VEL\_TEMP_{T_i-d}^i\Big\}_{d=1}^{12},\, \Big\{ACCEL\_TEMP_{T_i-d}^i\Big\}_{d=1}^{12} \Bigg)\, .
    \end{split}
    \end{equation}       
 
Therefore, the covariate vector consists of 81 variables and is represented as follows:
\begin{equation}
    \boldsymbol{x}_i=(\boldsymbol{z}_i,\boldsymbol{w}_i).
\end{equation}

Finally, our dataset $\mathcal{D}\in \mathbb{R}^{348\times 82}$ includes both cross-sectional and longitudinal variables that, as will be shown later, effectively characterize the corresponding agricultural yields. This dataset encompasses remote sensing variables related to climatic and geospatial conditions up to 12 weeks prior to the harvest. As will be discussed later, these temporal lags will enable us to identify causal relationships between these variables and agricultural yield. Next, having established the database for this study, we will outline the three methodologies we will employ to obtain our results and conclusions.

\subsubsection{Regression with Elastic-Net Regularization}

Given the dataset $\mathcal{D}$, our objective is to forecast agricultural yield $y_i$ using the predictors $\boldsymbol{x}_i$ defined earlier. To achieve this, we assume a linear regression structure between the variables:
\begin{equation}\label{lm}
    y_i=\beta_0 + \boldsymbol{x}_i^\top \beta + \xi_i,
\end{equation}
where the intercept $\beta_0$ and the weights $\beta=(\beta_1,\dots,\beta_{81})$ are unknown parameters, and $\xi_i$ represents the error term. Since our goal is to construct a simple and parsimonious model, and given that we have a large number of predictor variables that are closely related, we choose to induce moderate sparsity in the parameter vector. Therefore, to obtain the parameter estimates $(\hat{\beta}_0, \hat{\beta})$, we opt for Elastic-Net regularization \cite{zou2005regularization}, which involves solving the convex optimization problem:
\begin{equation}\label{elastic-net}
    \underset{(\beta_0, \beta) \in \mathbb{R} \times \mathbb{R}^{81}}{\text{min}} \left\{ \frac{1}{2} \sum_{i=1}^{N} \left( y_i - \beta_0 - \boldsymbol{x}_i^\top \beta \right)^2 + \lambda \left[ \frac{1}{2}(1 - \alpha)\|\beta\|_2^2 + \alpha \|\beta\|_1 \right] \right\},
\end{equation}
where $|\beta|p = \left( \sum{i=1}^{81} |\beta_i|^p \right)^{1/p}$ represents the standard $\ell^p$ norm. The penalty hyperparameter $\lambda \geq 0$ controls the complexity of the resulting model, while the hyperparameter $0 \leq \alpha \leq 1$ governs the desired level of sparsity. For more details, see \cite{hastie2015statistical}. We chose to set $\alpha=0.02$, which assigns more weight to the $\ell^2$ norm compared to the $\ell^1$ norm. This asymmetry in the weights results in a calibration of the induced sparsity in the parameter vector $\beta$ that aligns with qualitative field criteria regarding the expected relationships.

\begin{figure}[h!]
    \centering
    \includegraphics[width=0.9\linewidth]{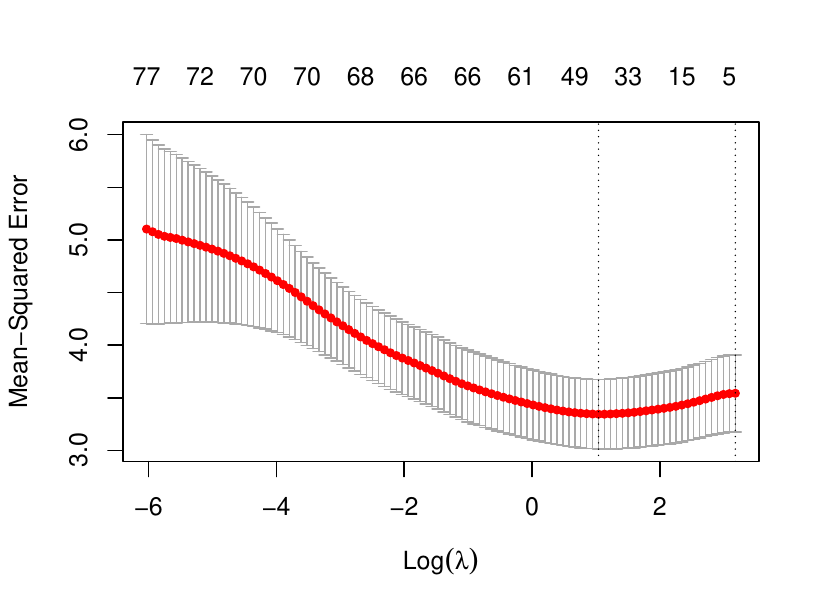}
    \caption{Graph to determine the optimal value of $\lambda$. Vertical axis: the MSE calculated through cross-validation. Horizontal axis: values of $\lambda$ on a logarithmic scale. The red line represents the average value of the MSE, and the gray bands represent their respective confidence intervals.}
    \label{Res: Lambda}
\end{figure}

To determine the optimal value of $\lambda$, we employed the standard cross-validation criterion, evaluating the mean squared error (MSE) of the cross-validation for different values of $\lambda$ on a logarithmic scale, as shown in Figure \boldref{Res: Lambda}. This procedure yielded an optimal value of $\lambda=2.58$. Finally, once both hyperparameters were calibrated, we solved the problem \eqref{elastic-net} using convex optimization algorithms extensively detailed in \cite{hastie2015statistical}. Consequently, the solution to \eqref{elastic-net} provides us with a sparse estimate of the parameter vector for the model \eqref{lm}.

\subsubsection{Gradient Tree Boosting}\label{sec:XGBOOST}
Let $q:\mathbb{R}^{81}\rightarrow \mathcal{T}$ represent the structure of a tree that maps the characteristics of a crop $\boldsymbol{x}_i$ to the index of the corresponding leaf. The weight vector of its leaves is given by $\omega=(\omega_1,\dots,\omega{|\mathcal{T}|})\in \mathbb{R}^{|\mathcal{T}|}$, where $\omega_k$ denotes the score of the $k$-th leaf. Here, $\mathcal{T}$ is the set of leaves of the tree, and $|\mathcal{T}|$ indicates the total number of leaves. To obtain the prediction of agricultural yield $\hat{y}_i$, we will use an additive ensemble $\kappa$ of these trees, denoted by $\phi$, which can be expressed as follows:
\begin{equation}\label{boosting}
 \hat{y}_i = \phi(\boldsymbol{x}_i) = \sum_{k=1}^{\kappa} f_k(\boldsymbol{x}_i), \quad f_k \in \mathcal{F},   
\end{equation}
where $\mathcal{F} = \big\{f:\mathbb{R}^{81}\rightarrow \mathbb{R} \mid f(\boldsymbol{x}_i) = \omega_{q(\boldsymbol{x}_i)}\big\}$ denotes the space of regression trees, also known as CART \cite{breiman2017classification}. It is important to note that each $f_k$ corresponds to an independent tree structure $q$ with its associated leaf weights $\omega$.

\begin{figure}[h!]
    \centering
    \includegraphics[width=0.85\linewidth]{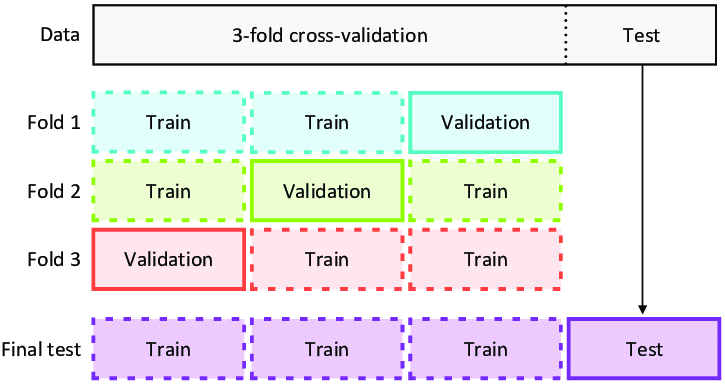}
    \caption{This is a schematic representation of 3-fold cross-validation, where in each iteration, one fold is used for validation while the other two are used for training. The process is repeated three times to identify the optimal combination of hyperparameters. For more details, see \cite{duran2020prometeo}.}
    \label{fig: CrossValidation}
\end{figure}

Based on the dataset $\mathcal{D}$, the learning of the functions $f_k$ used in the model \eqref{boosting} is achieved by solving the regularized optimization problem:
\begin{equation}\label{opt-boosting}
\underset{f_k\in \mathcal{F},\,\forall k}{\text{min}}\, \sum_{i=1}^{N} (\hat{y}_i- y_i)^2 + \sum_{k=1}^{\kappa}\bigg\{ \gamma |\mathcal{T}| + \frac{1}{2} \lambda \|\omega\|^2_2\bigg\},   
\end{equation}
where the hyperparameter $\gamma$ penalizes complexity due to the depth of the trees, and $\lambda$ regularizes the weights of the trees to prevent overfitting. The standard learning algorithm used to solve \eqref{opt-boosting} is based on gradient methods, which is why this model is commonly referred to as Gradient Tree Boosting (XGBoost) \cite{chen2016xgboost}. The hyperparameters in \eqref{opt-boosting} are calibrated using established methodologies. Specifically, we perform optimal selection through 3-fold cross-validation, as illustrated in Figure \boldref{fig: CrossValidation}. Through this procedure, we obtain optimal values of $\gamma=0.1$ and $\lambda=0.6$. Additionally, to regularize the set of tree leaves $\mathcal{T}$, we optimally limit the maximum depth of the trees to 5.
\subsubsection{Semi-parametric Additive Model}
As a semi-parametric alternative, we choose a Generalized Additive Model (GAM) structure, which can be expressed in our case as follows:
\begin{equation}\label{gam}
\hat{y}_i=\theta_0+\boldsymbol{z}_i^\top\theta+\sum_{j=1}^{72} f_j(\boldsymbol{w}_i^{(j)}),
\end{equation}
where $\boldsymbol{w}_i^{(j)}$ is the $j$-th coordinate of the vector defined in \eqref{w}. In this model, $\theta \in \mathbb{R}^9$ represents a parameter vector, $\theta_0 \in \mathbb{R}$ is the intercept, and $f_j$ are smooth functions to be estimated using the dataset $\mathcal{D}$. To estimate the model in \eqref{gam}, we adopt the widely used approach of representing the functions $f_j$ with reduced-rank smoothing splines that result from solving variational problems. For more details, see \cite{wood2017generalized}.
\section{Results and discussion}\label{sec:3}

    As mentioned earlier, we have two groups of predictor variables. The first group, $\boldsymbol{z}_i$, consists of variables that characterize the use of good agricultural practices in crops. These variables serve as control variables; that is, we are not interested in their direct effects but rather use them to control for the influences of other factors that may affect the relationship between the predictors and the response variable. The second group, $\boldsymbol{w}_i$, includes the velocities and accelerations of the remote sensing variables—specifically, the velocities and accelerations of NDVI, PREC, TEMP, and their respective time lags, as defined in \eqref{vel_NDVI}, \eqref{acel_NDVI}, and \eqref{w}. Our primary interest in this study is to understand the causal relationships between the remote sensing variables and agricultural yield, particularly focusing on the parameters (or coefficients) associated with the variables in the vector $\boldsymbol{w}_i$.
        
\begin{figure}[h!]
            \centering
            \includegraphics[width=0.8\linewidth]{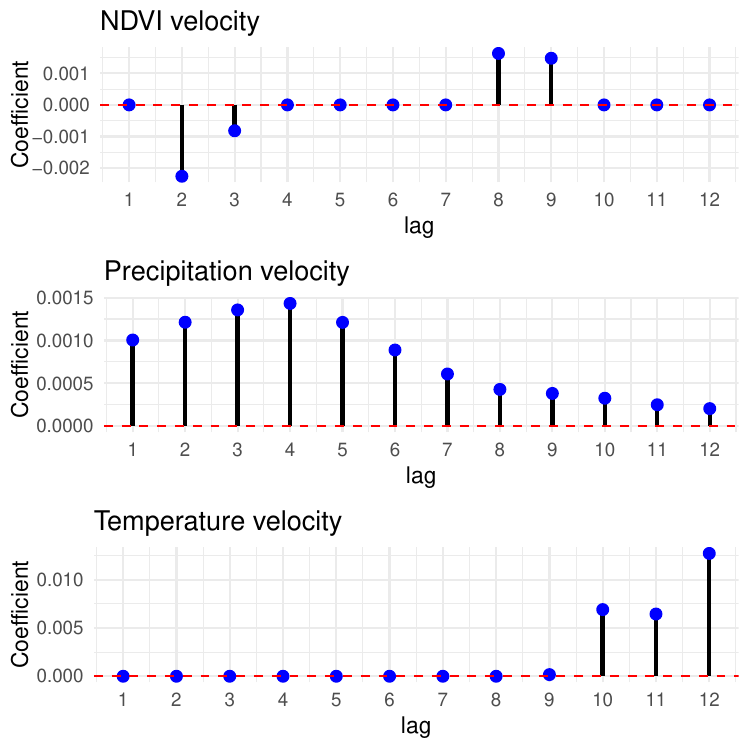}
            \caption{Representation of the velocity variables. The chart features three subplots, each illustrating the relationship between a coefficient and the lag for different variables. Additionally, each subplot includes bars that indicate the direction and effect of the lag variable.}
            \label{fig: Velocidad}
        \end{figure}

    Regarding the velocity variables, the parameter vectors obtained from model \eqref{lm} through \eqref{elastic-net} for NDVI and TEMP are highly sparse, in contrast to the parameter vector for PREC, which is dense; see Figure \boldref{fig: Velocidad}. The sparsity in the velocities of NDVI and Temperature suggests that the effects of variations in these variables on agricultural yield are delayed. For instance, first-order variations in NDVI affect agricultural yield only 8 or 9 weeks after they occur, as shown in Figure \boldref{fig: Velocidad}. A similar pattern is observed with Temperature variations, which have a lag of 10 to 12 weeks. It is important to note that these lag periods are not precise and may vary with the sample. Nonetheless, a significant qualitative conclusion is that variations in these two variables do not immediately impact agricultural yield. The lag associated with both climatic variables may be attributed to the crop germination process.
        
\begin{figure}[h!]
            \centering
            \includegraphics[width=0.8\linewidth]{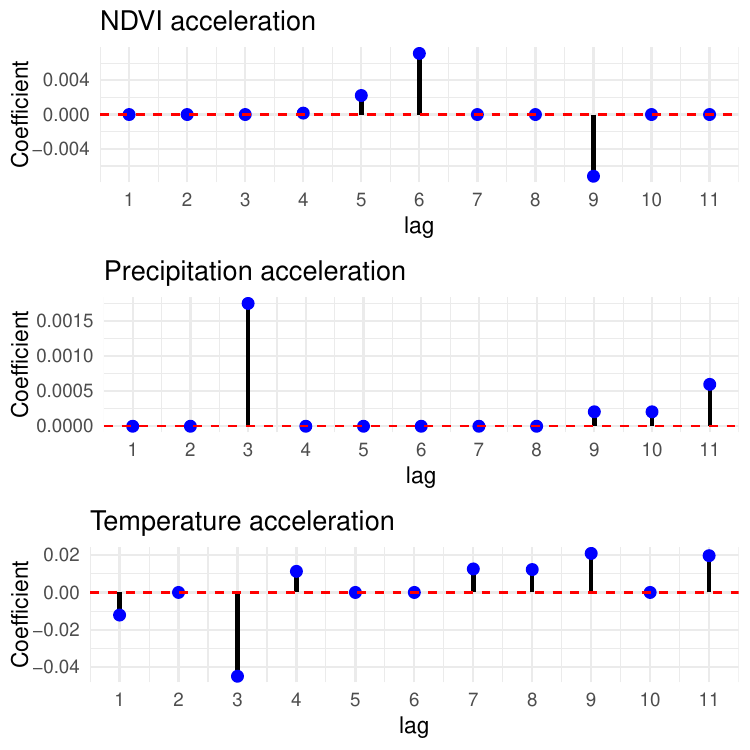}
            \caption{Representation of the acceleration variables. The chart features three subplots, each illustrating the relationship between a coefficient and the lag for various variables. Additionally, each subplot includes bars that indicate both the direction and effect of the lag variable.}
            \label{fig: aceleracion}
        \end{figure}

    In contrast, variations in Precipitation have an immediate and lasting effect on agricultural yield. The scenario changes when considering the acceleration variables: second-order variations in NDVI impact agricultural yield between 6 to 9 weeks later, while similar variations in Precipitation and Temperature affect it approximately 3 weeks later; see Figure \boldref{fig: aceleracion}. The acceleration parameters associated with all three variables, obtained from model \eqref{lm} through \eqref{elastic-net}, are also sparse vectors.

Based on these analyses, we can assert that the velocities and accelerations of NDVI, Precipitation, and Temperature have a causal effect on agricultural yield. Since these causal relationships involve time lags, we can state that the relationship is in the Granger sense \cite{granger1969investigating}. It is essential to highlight that these causal relationships are absent in the original remote sensing variables; constructing the velocity and acceleration variables is necessary to establish such causal connections.

To compare predictive capacity following the identification of causal relationships between the remote sensing variables and agricultural yield, we chose to use the XGBoost model described in \eqref{boosting} and \eqref{opt-boosting}. This approach allows us to capture the complex non-linear patterns between the predictor variables $\boldsymbol{x}_i$ and the response variable $y_i$. Given that XGBoost is a highly flexible non-parametric model, combined with the constraints posed by a small sample size, there is a risk of overfitting when relationships are spurious or synthetic. In this context, our construction of velocity and acceleration variables helps mitigate this risk, as these variables maintain causal relationships with our response variable.

Since the Elastic-Net regularized regression model is fully parametric and XGBoost is entirely non-parametric, we will also include a semi-parametric alternative in our comparative prediction analysis: the GAM model described in \eqref{gam}. To compare the three models, the dataset was split into training and test sets in an 80\% to 20\% ratio, respectively. The performance of each model in both samples is presented in Table \boldref{tab:mse-modelos}.

\begin{table}[H]
\centering
\begin{tabular}{lcc}
\hline
\textbf{Model} & \textbf{Train} & \textbf{Validation} \\ \hline
Elastic-Net Regularization    & 2.81           & 3.93               \\
XGBoost  & 0.84           & 2.94               \\
GAM      & 2.29           & 4.09               \\ \hline
\end{tabular}
\caption{Mean Square Error (MSE) results for the three models applied to training and test data.}
\label{tab:mse-modelos}
\end{table}

The XGBoost model demonstrates a good fit, as evidenced by its low MSE value; however, the notable difference in MSE between the training and test samples suggests potential overfitting. In contrast, the Elastic-Net regularized regression model shows a smaller increase in MSE values, indicating greater stability and better generalization ability. Meanwhile, the GAM model exhibits high MSE values in both the training and test samples compared to the other two models.

These conclusions regarding the performance of the three models may be attributed to the limited amount of data available for this study, which restricts XGBoost's ability to perform effective cross-validation and adequately capture the interactions between the variables and agricultural yield. The most significant finding is that the construction of velocity and acceleration variables derived from remote sensing data, due to their causal nature, can substantially enhance the development of models for predicting agricultural yields.

\section{Conclusions}\label{sec:4}
This study demonstrates that remote sensing variables contain valuable predictive information about agricultural production. However, the relationships between these variables and production are neither linear nor straightforward; instead, they are complex and difficult to identify without appropriate dynamic transformations. For this reason, we propose employing sparsity techniques based on Elastic-Net regularized regression, alongside dynamic lag criteria. This approach enables effective management of the correlations between the velocity and acceleration variables derived from remote sensing data, while also enhancing the accuracy of causal relationship identification.

Furthermore, the results indicate a causal relationship, in the Granger sense, between the remote sensing variables and agricultural yield, highlighting their predictive capability—particularly when integrated with agricultural census data. Our main contribution lies in identifying the types of dynamic transformations that should be applied to remote sensing variables for effective use in agricultural prediction models. Therefore, utilizing these transformations along with machine learning techniques represents a promising strategy for developing simpler and more accurate predictive models, significantly improving forecasting capabilities in agriculture.

\section{Contributions and Findings}\label{sec:5}

This work makes several significant contributions. First, it emphasizes the integration of remote sensing data with agricultural censuses, resulting in the creation of a robust and enhanced database. This combination provides accurate and up-to-date information on both the study area and the crop, facilitating more comprehensive analyses. Second, variable engineering was conducted based on the extracted agricultural and climatic data, enabling the capture of non-linear patterns that influence crop yield. Finally, the results and conclusions of this study lead to substantial improvements in crop forecasting, offering a deeper understanding of how factors such as climate, soil moisture, and vegetation health impact growth and yield.


\end{document}